\documentclass[10pt,twoside]{article}
\usepackage{amsmath}
\usepackage[latin1]{inputenc}

\numberwithin{equation}{section}

\begin{document}

\title{Reparameterization of the scalar potential \\
in the two Higgs doublet model}
\author{Rodolfo A. Diaz\thanks{%
radiazs@unal.edu.co}, Viviana Dionicio\thanks{lvdioniciol@unal.edu.co}, R. Martinez\thanks{%
remartinezm@unal.edu.co} \\
Universidad Nacional de Colombia, \\
Departamento de Física. Bogotá, Colombia.}
\date{}
\maketitle
\vspace{-1cm}
\begin{abstract}
In the two Higgs doublet model with no additional symmetries in the scalar
sector (different from the gauge and Lorentz symmetries), it is customary to
reparameterize the model by rotating the scalar doublets such that one of
the vacuum expectation values vanishes. It is well known that the Yukawa
sector of the model is unaffected by such transformation. Notwithstanding,
since the Higgs potential must also be transformed, it is necessary to show
that such sector is also unaltered in its physical content. We
demostrate that the physical content of the potential is invariant even when
the charge conjugation symmetry is demanded.

PACS \{12.60.Fr, 11.30.Er, 11.15.Ex, 11.30.Hv\}

\end{abstract}

\vspace{-0.7cm}

\section{Introduction}

One of the simplest extension of the SM is the so called two Higgs doublet
model (2HDM), which consists of adding a new doublet with the same quantum
numbers as the former. The most studied of this kind of models are the so
called 2HDM type I and 2HDM type II, according to the discrete symmetry
imposed in the scalar sector that provides different Yukawa couplings in the
up and down sectors. In the model of type I, only one Higgs doublet gives
mass to the up and down sectors; while in the model of type II one doublet
gives mass to the up type quarks while the other doublet provides the mass
to the down type quarks. Since each doublet could acquire a vacuum
expectation value (VEV), the presence of two different VEV's enables us to
explain the hierarchy mass problem in the third generation of quarks. In the
case in which no discrete symmetry is introduced the two doublets can couple
to both types of quarks, generating flavor changing neutral currents (FCNC);
because in this case the mass matrix is generated by two different matrices
which cannot be diagonalized at the same time. In this case we talk about
2HDM type III. In the case of the 2HDM III, it is possible to make a
rotation between the two higgs doublets which is equivalent to choose a new
basis, and the physical content of the Lagrangian remains invariant. This
fact permits to make a rotation such that one of the VEV's vanishes.

In general the 2HDM has two CP-even neutral scalar Higgs bosons $%
(h^{0},H^{0})$ which are mixed through an angle denoted as $\alpha $; two
CP-odd scalar Fields $(A^{0},G_{Z}^{0})$ where the first correspond to a
physical particle and the second is a Goldstone boson associated with $%
Z_{\mu }$; and finally, two types of charged fields $H^{\pm },G_{W}^{\pm }$
which correspond to two scalar charged particles and two would be Goldstone
bosons associated with $W^{\pm }$. They are mixed through an angle denoted
by $\beta $ where $\tan \beta =v_{1}/v_{2}$ with $v_{1}$ $v_{2}$ denoting
the two VEV's. However, in the model where one of the VEV is taken out by
the rotation, the would be Goldstone bosons and scalar bosons do not mix
because they belong to different doublets due to the redefinition of the
coordinate system for the doublets. For $SU(N)$ gauge group with $m-$%
multiplet fundamental representation which get VEV $<H_{i}>=v_{i}$ it is
always possible to rotate the system such that $<H_{1}>=v_{1}$ and $%
<H_{i}>=0 $ with $i=2...m$ \cite{SherGeorgi}.

In summary, in the absence of additional symmetries (different from the
gauge and Lorentz symmetry) in the scalar sector, we can always choose a
basis in which only one doublet acquire a vacuum expectation value (VEV)
without affecting the physical meaning of the model \cite{SherGeorgi}. This
change of basis consists of a unitary transformation between the two Higgs
doublets that cancels one of the VEV. Nevertheless, the literature usually
study this rotation in the Yukawa sector only. However, even in the case in
which no additional symmetry is taken in the Yukawa sector, charge
conjugation invariance ($C-$invariance) is assumed in the Higgs potential 
\cite{Hunter, Santos}, and since the same choice of basis must be done in
this sector, it is necessary to check that a change of basis keeps the Higgs
potential unaltered in its physical content. The purpose of the present
paper is to show explicitly that the potential maintains its physical
meaning even when $C-$invariance is imposed as long as no additional
symmetries in the potential are demanded.

\section{Rotation of the Yukawa Lagrangian in the 2HDM type III\label%
{rotation Yuk}}

We start defining two Higgs doublets with VEV's 
\begin{equation}
\Phi _{1,2}^{\prime }=\left( 
\begin{array}{c}
\left( \phi _{1,2}^{+}\right) ^{\prime } \\ 
\left( \phi _{1,2}^{0}\right) ^{\prime }%
\end{array}%
\right) =\left( 
\begin{array}{c}
\left( \phi _{1,2}^{+}\right) ^{\prime } \\ 
\left( h_{1}+v_{1}+ig_{1}\right) ^{\prime }%
\end{array}%
\right) \;;\;\widetilde{\Phi }_{1,2}^{\prime }=i\sigma _{2}\Phi
_{1,2}^{\prime }\;;\;\langle \Phi _{1,2}^{\prime }\rangle =v_{1,2}^{\prime }
\label{doubletnotri}
\end{equation}%
and writing the Yukawa Lagrangian of the most general two Higgs doublet
model (the so called 2HDM type III) 
\begin{eqnarray}
-£_{Y} &=&\widetilde{\eta }_{ij}^{U,0}\overline{Q}_{iL}^{0}\widetilde{\Phi }%
_{1}^{\prime }U_{jR}^{0}+\widetilde{\eta }_{ij}^{D,0}\overline{Q}%
_{iL}^{0}\Phi _{1}^{\prime }D_{jR}^{0}+\widetilde{\xi }_{ij}^{U,0}\overline{Q%
}_{iL}^{0}\widetilde{\Phi }_{2}^{\prime }U_{jR}^{0}+\widetilde{\xi }%
_{ij}^{D,0}\overline{Q}_{iL}^{0}\Phi _{2}^{\prime }D_{jR}^{0}  \notag \\
&&+\text{lepton sector}+h.c.  \label{Yuknotri}
\end{eqnarray}%
where $Q_{iL}^{0}$ denotes the left-handed quark doublets with $i$ the
family index, $U_{jR}^{0}$,$\ D_{jR}^{0}$ correspond to the right-handed
singlets of up-type and down-type quarks respectively. The superscript
\textquotedblleft $0$\textquotedblright\ means that we are dealing with
gauge eigenstates. Finally, $\widetilde{\eta }_{ij}^{U,0},\ \widetilde{\xi }%
_{ij}^{D,0}$ correspond to the Yukawa vertices giving in general a mixing
among families. We can make a rotation between the doublets 
\begin{equation}
\left( 
\begin{array}{c}
\Phi _{1} \\ 
\Phi _{2}%
\end{array}%
\right) \equiv \left( 
\begin{array}{cc}
\cos \theta & \sin \theta \\ 
-\sin \theta & \cos \theta%
\end{array}%
\right) \left( 
\begin{array}{c}
\Phi _{1}^{\prime } \\ 
\Phi _{2}^{\prime }%
\end{array}%
\right)  \label{rotdoublet}
\end{equation}%
and defined some new rotated Yukawa couplings%
\begin{equation}
\left( 
\begin{array}{c}
\eta _{ij}^{\left( U,D\right) ,0} \\ 
\xi _{ij}^{\left( U,D\right) ,0}%
\end{array}%
\right) =\left( 
\begin{array}{cc}
\cos \theta & \sin \theta \\ 
-\sin \theta & \cos \theta%
\end{array}%
\right) \left( 
\begin{array}{c}
\widetilde{\eta }_{ij}^{\left( U,D\right) ,0} \\ 
\widetilde{\xi }_{ij}^{\left( U,D\right) ,0}%
\end{array}%
\right)  \label{rotmat}
\end{equation}

We shall deal with the quark sector only since the results for the lepton
sector will be straightforward. In terms of $\Phi _{1},\;\Phi _{2}$,\ $\eta
_{ij}^{\left( U,D\right) ,0}$ and $\xi _{ij}^{\left( U,D\right) ,0\ }$the
Yukawa Lagrangian could be rewritten as 
\begin{equation*}
-£_{Y}=\overline{Q}_{iL}^{0}\eta _{ij}^{U,0}\widetilde{\Phi }_{1}U_{jR}^{0}+%
\overline{Q}_{iL}^{0}\eta _{ij}^{D,0}\Phi _{1}D_{jR}^{0}+\overline{Q}%
_{iL}^{0}\xi _{ij}^{U,0}\widetilde{\Phi }_{2}U_{jR}^{0}+\overline{Q}%
_{iL}^{0}\xi _{ij}^{D,0}\Phi _{2}D_{jR}^{0}+h.c.
\end{equation*}%
with the same form as the original Lagrangian if we forget the prime
notation. Consequently, the combined rotations (\ref{rotdoublet}, \ref%
{rotmat}) do not have physical consequences since it is basically a change
of basis. In particular we can choose $\theta =\beta \;$where $\tan \beta
\equiv v_{2}^{\prime }/v_{1}^{\prime }\ $such that

\begin{eqnarray}
\langle \Phi _{1}\rangle &=&\cos \beta \langle \Phi _{1}^{\prime }\rangle
+\sin \beta \langle \Phi _{2}^{\prime }\rangle =\sqrt{v_{1}^{\prime
2}+v_{2}^{\prime 2}}\equiv v  \notag \\
\langle \Phi _{2}\rangle &=&-\sin \beta \langle \Phi _{1}^{\prime }\rangle
+\cos \beta \langle \Phi _{2}^{\prime }\rangle =0  \label{VEV zero}
\end{eqnarray}%
In whose case we managed to get $\langle \Phi _{2}\rangle =0$.\ Since this
lagrangian contains exactly the same physical information as the first one,
we conclude that in model type III the parameter $\tan \beta \;$is totally
spurious, and we can assume without any loss of generality that one of the
VEV is zero.

On the other hand, it is possible to reverse the steps above and start from
the representation in which $\langle \Phi _{2}\rangle =0\;$(the
\textquotedblleft fundamental representation\textquotedblright ) and make a
rotation of the Higgs doublets from which the $\tan \beta \;$parameter
arises. Although these rotations provides an spurious parameter ($\tan \beta 
$), they have the advantage of making the comparison between the model type
II with the model type III more apparent \cite{PRDbasis}, and same for the
comparison between model type I and type III. It is important to emphasize
that the mixing matrices$\ \eta _{ij}^{\left( U,D\right) ,0},\xi
_{ij}^{\left( U,D\right) ,0}$ depend explicitly on the $\tan \beta $
parameter, thus, they are basis dependent. Nevertheless, it can be
explicitly shown that the whole couplings are basis independent \cite%
{tesisRod} as expected.

We should emphasize that the invariance of Lagrangian (\ref{Yuknotri})\
under rotation (\ref{rotdoublet}) requires no additional symmetries to be
imposed in such Lagrangian. Notwithstanding, even when Lagrangian (\ref%
{Yuknotri}) is leaving in the most general form it is customary to impose a $%
C-$invariance in the Higgs potential, and since rotation (\ref{rotdoublet})
must be applied to such potential as well, we should check that the
transformation described by Eq. (\ref{rotdoublet}) still leave the potential
unaltered in its physical content.

\section{Rotation in the Higgs potential\label{rotation pot}}

After examining the rotation in the Yukawa sector, we shall see how this
transformation changes the parameters in the potential, and show that the
physical content of the potential remains intact. For which we should
examine the physical parameters of the model and show that they are basis
invariant.

\subsection{Transformation of the parameters in the potential}

Let us start from an arbitrary parameterization in which both VEV are in
general different from zero. The most general renormalizable and gauge
invariant potential read 
\begin{eqnarray}
V_{g} &=&-\widetilde{\mu }_{1}^{2}\widehat{A}^{\prime }-\widetilde{\mu }%
_{2}^{2}\widehat{B}^{\prime }-\widetilde{\mu }_{3}^{2}\widehat{C}^{\prime }-%
\widetilde{\mu }_{4}^{2}\widehat{D}^{\prime }+\widetilde{\lambda }_{1}%
\widehat{A}^{\prime 2}+\widetilde{\lambda }_{2}\widehat{B}^{\prime 2}+%
\widetilde{\lambda }_{3}\widehat{C}^{\prime 2}+\widetilde{\lambda }_{4}%
\widehat{D}^{\prime 2}  \notag \\
&&+\widetilde{\lambda }_{5}\widehat{A}^{\prime }\widehat{B}^{\prime }+%
\widetilde{\lambda }_{6}\widehat{A}^{\prime }\widehat{C}^{\prime }+%
\widetilde{\lambda }_{8}\widehat{A}^{\prime }\widehat{D}^{\prime }+%
\widetilde{\lambda }_{7}\widehat{B}^{\prime }\widehat{C}^{\prime }+%
\widetilde{\lambda }_{9}\widehat{B}^{\prime }\widehat{D}^{\prime }+%
\widetilde{\lambda }_{10}\widehat{C}^{\prime }\widehat{D}^{\prime }
\label{Vg2}
\end{eqnarray}%
where we have defined four independent gauge invariant hermitian operators 
\begin{eqnarray*}
\widehat{A}^{\prime } &\equiv &\Phi _{1}^{\prime \dagger }\Phi _{1}^{\prime
}\;,\;\widehat{B}^{\prime }\equiv \Phi _{2}^{\prime \dagger }\Phi
_{2}^{\prime },\;\widehat{C}^{\prime }\equiv \frac{1}{2}\left( \Phi
_{1}^{\prime \dagger }\Phi _{2}^{\prime }+\Phi _{2}^{\prime \dagger }\Phi
_{1}^{\prime }\right) =\text{Re}\left( \Phi _{1}^{\prime \dagger }\Phi
_{2}^{\prime }\right) ,\; \\
\widehat{D}^{\prime } &\equiv &-\frac{i}{2}\left( \Phi _{1}^{\prime \dagger
}\Phi _{2}^{\prime }-\Phi _{2}^{\prime \dagger }\Phi _{1}^{\prime }\right) =%
\text{Im}\left( \Phi _{1}^{\prime \dagger }\Phi _{2}^{\prime }\right)
\end{eqnarray*}%
In section \ref{rotation Yuk} we showed that the rotation described by Eq. (%
\ref{rotdoublet}) can be done without changing the physical content of the
Yukawa Lagrangian. In order to show the invariance of the physical content
in the scalar potential, we shall calculate the way in which the $\widetilde{%
\mu }_{i},\,\widetilde{\lambda }_{i}\;$parameters transform under this
rotation. First, we calculate the way in which the operators $\widehat{A}%
^{\prime },\;\widehat{B}^{\prime },\;\widehat{C}^{\prime },\;\widehat{D}%
^{\prime }$ transform. Taking into account Eq. (\ref{rotdoublet}) we get 
\begin{eqnarray*}
\widehat{A}^{\prime } &\equiv &\Phi _{1}^{\prime \dagger }\Phi _{1}^{\prime
}=\left( \Phi _{1}^{\dagger }\cos \theta -\Phi _{2}^{\dagger }\sin \theta
\right) \left( \Phi _{1}\cos \theta -\Phi _{2}\sin \theta \right) \\
&=&\Phi _{1}^{\dagger }\Phi _{1}\cos ^{2}\theta -2\cos \theta \sin \theta
\left( \frac{\Phi _{1}^{\dagger }\Phi _{2}+\Phi _{2}^{\dagger }\Phi _{1}}{2}%
\right) +\Phi _{2}^{\dagger }\Phi _{2}\sin ^{2}\theta \\
&=&\widehat{A}\cos ^{2}\theta +\widehat{B}\sin ^{2}\theta -\sin 2\theta 
\widehat{C}
\end{eqnarray*}%
we obtain the transformations for the other operators in a similar way, the
results read 
\begin{eqnarray*}
\widehat{A}^{\prime } &=&\widehat{A}\cos ^{2}\theta +\widehat{B}\sin
^{2}\theta -\widehat{C}\sin 2\theta \\
\widehat{B}^{\prime } &=&\widehat{A}\sin ^{2}\theta +\widehat{B}\cos
^{2}\theta +\widehat{C}\sin 2\theta \\
\widehat{C}^{\prime } &=&\frac{1}{2}\widehat{A}\sin 2\theta -\frac{1}{2}%
\widehat{B}\sin 2\theta +\widehat{C}\cos 2\theta \\
\widehat{D}^{\prime } &=&\widehat{D}
\end{eqnarray*}%
\begin{eqnarray*}
\widehat{A}^{\prime 2} &=&\widehat{A}^{2}\cos ^{4}\theta +\widehat{B}%
^{2}\sin ^{4}\theta +\widehat{C}^{2}\sin ^{2}2\theta +\frac{1}{2}\widehat{A}%
\widehat{B}\sin ^{2}2\theta -2\widehat{A}\widehat{C}\sin 2\theta \cos
^{2}\theta -2\widehat{B}\widehat{C}\sin ^{2}\theta \sin 2\theta \\
\widehat{B}^{\prime 2} &=&\widehat{A}^{2}\sin ^{4}\theta +\widehat{B}%
^{2}\cos ^{4}\theta +\widehat{C}^{2}\sin ^{2}2\theta +\frac{1}{2}\widehat{A}%
\widehat{B}\sin ^{2}2\theta +2\widehat{A}\widehat{C}\sin 2\theta \sin
^{2}\theta +2\widehat{B}\widehat{C}\sin 2\theta \cos ^{2}\theta \\
\widehat{C}^{\prime 2} &=&\frac{1}{4}\left( \widehat{A}^{2}+\widehat{B}%
^{2}\right) \sin ^{2}2\theta +\widehat{C}^{2}\cos ^{2}2\theta -\frac{1}{2}%
\widehat{A}\widehat{B}\sin ^{2}2\theta +\frac{1}{2}\widehat{A}\widehat{C}%
\sin 4\theta -\frac{1}{2}\widehat{B}\widehat{C}\sin 4\theta \\
\widehat{D}^{\prime 2} &=&\widehat{D}^{2}
\end{eqnarray*}%
\begin{eqnarray*}
\widehat{A}^{\prime }\widehat{B}^{\prime } &=&\left( \frac{1}{4}\widehat{A}%
^{2}+\frac{1}{4}\widehat{B}^{2}-\widehat{C}^{2}\right) \sin ^{2}2\theta +%
\widehat{A}\widehat{B}\left( \cos ^{4}\theta +\sin ^{4}\theta \right)
+\left( \widehat{A}\widehat{C}-\widehat{B}\widehat{C}\right) \sin 2\theta
\cos 2\theta \\
\widehat{A}^{\prime }\widehat{C}^{\prime } &=&\frac{1}{2}\widehat{A}^{2}\sin
2\theta \cos ^{2}\theta -\frac{1}{2}\widehat{B}^{2}\sin ^{2}\theta \sin
2\theta -\widehat{C}^{2}\sin 2\theta \cos 2\theta -\frac{1}{4}\widehat{A}%
\widehat{B}\sin 4\theta \\
&&+\widehat{A}\widehat{C}\left( 4\cos ^{2}\theta -3\right) \cos ^{2}\theta +%
\widehat{B}\widehat{C}\left( 4\cos ^{2}\theta -1\right) \sin ^{2}\theta \\
\widehat{A}^{\prime }\widehat{D}^{\prime } &=&\widehat{A}\widehat{D}\cos
^{2}\theta +\widehat{B}\widehat{D}\sin ^{2}\theta -\widehat{C}\widehat{D}%
\sin 2\theta
\end{eqnarray*}%
\begin{eqnarray}
\widehat{B}^{\prime }\widehat{C}^{\prime } &=&\frac{1}{2}\widehat{A}^{2}\sin
2\theta \sin ^{2}\theta -\frac{1}{2}\widehat{B}^{2}\sin 2\theta \cos
^{2}\theta +\frac{1}{2}\widehat{C}^{2}\sin 4\theta +\frac{1}{4}\widehat{A}%
\widehat{B}\sin 4\theta  \notag \\
&&+\widehat{A}\widehat{C}\left( \cos 2\theta +2\cos ^{2}\theta \right) \sin
^{2}\theta +\widehat{B}\widehat{C}\left( \cos 2\theta -2\sin ^{2}\theta
\right) \cos ^{2}\theta  \notag \\
\widehat{B}^{\prime }\widehat{D}^{\prime } &=&\widehat{A}\widehat{D}\sin
^{2}\theta +\widehat{B}\widehat{D}\cos ^{2}\theta +\widehat{C}\widehat{D}%
\sin 2\theta  \notag \\
\widehat{C}^{\prime }\widehat{D}^{\prime } &=&\frac{1}{2}\left( \widehat{A}%
\widehat{D}-\widehat{B}\widehat{D}\right) \sin 2\theta +\widehat{C}\widehat{D%
}\cos 2\theta  \label{operator transf}
\end{eqnarray}

Now, we can build up a new parameterization of the potential such that 
\begin{eqnarray}
V_{g} &=&-\mu _{1}^{2}\widehat{A}-\mu _{2}^{2}\widehat{B}-\mu _{3}^{2}%
\widehat{C}-\mu _{4}^{2}\widehat{D}+\lambda _{1}\widehat{A}^{2}+\lambda _{2}%
\widehat{B}^{2}+\lambda _{3}\widehat{C}^{2}+\lambda _{4}\widehat{D}^{2} 
\notag \\
&&+\lambda _{5}\widehat{A}\widehat{B}+\lambda _{6}\widehat{A}\widehat{C}%
+\lambda _{8}\widehat{A}\widehat{D}+\lambda _{7}\widehat{B}\widehat{C}%
+\lambda _{9}\widehat{B}\widehat{D}+\lambda _{10}\widehat{C}\widehat{D}
\label{Vg rotated}
\end{eqnarray}%
in order to find the values of $\mu _{i},\;\lambda _{i}\;$in terms of $%
\widetilde{\mu }_{i},\;\widetilde{\lambda }_{i}$, we use the Eqs. (\ref{Vg2}%
), and (\ref{operator transf}) to write e.g. the coefficient proportional to
the operator $\widehat{A}$, and these terms are compared with the term
proportional to the operator $\widehat{A}\;$in Eq. (\ref{Vg rotated})
obtaining 
\begin{equation*}
-\mu _{1}^{2}\widehat{A}=\left( -\widetilde{\mu }_{1}^{2}\cos ^{2}\theta -%
\widetilde{\mu }_{2}^{2}\sin ^{2}\theta -\widetilde{\mu }_{3}^{2}\sin \theta
\cos \theta \right) \widehat{A}
\end{equation*}%
therefore, the coefficient $\mu _{1}^{2}\;$is related to the parameters $%
\widetilde{\mu }_{i},\;\widetilde{\lambda }_{i}\;$in the following way 
\begin{equation*}
\mu _{1}^{2}=\left( \widetilde{\mu }_{1}^{2}\cos ^{2}\theta +\widetilde{\mu }%
_{2}^{2}\sin ^{2}\theta +\frac{1}{2}\widetilde{\mu }_{3}^{2}\sin 2\theta
\right)
\end{equation*}%
by the same token, the other sets$\;$of $\mu _{i},\;\lambda _{i}$ parameters
are related to the $\widetilde{\mu }_{i},\;\widetilde{\lambda }_{i}\;$%
parameters in the following way 
\begin{eqnarray*}
\mu _{1}^{2} &=&\left( \widetilde{\mu }_{1}^{2}\cos ^{2}\theta +\widetilde{%
\mu }_{2}^{2}\sin ^{2}\theta +\frac{1}{2}\widetilde{\mu }_{3}^{2}\sin
2\theta \right) \ \ ;\ \ \mu _{2}^{2}=\left( \widetilde{\mu }_{1}^{2}\sin
^{2}\theta +\widetilde{\mu }_{2}^{2}\cos ^{2}\theta -\frac{1}{2}\widetilde{%
\mu }_{3}^{2}\sin 2\theta \right) \\
\mu _{3}^{2} &=&\left( -\widetilde{\mu }_{1}^{2}\sin 2\theta +\widetilde{\mu 
}_{2}^{2}\sin 2\theta +\widetilde{\mu }_{3}^{2}\cos 2\theta \right) \ \ ;\ \
\mu _{4}^{2}=\widetilde{\mu }_{4}^{2}
\end{eqnarray*}

\begin{eqnarray*}
\lambda _{1} &=&\widetilde{\lambda }_{1}\cos ^{4}\theta +\widetilde{\lambda }%
_{2}\sin ^{4}\theta +\frac{1}{4}\left( \widetilde{\lambda }_{3}+\widetilde{%
\lambda }_{5}\right) \sin ^{2}2\theta +\frac{1}{2}\left( \widetilde{\lambda }%
_{6}\cos ^{2}\theta +\widetilde{\lambda }_{7}\sin ^{2}\theta \right) \sin
2\theta \\
\lambda _{2} &=&\widetilde{\lambda }_{1}\sin ^{4}\theta +\widetilde{\lambda }%
_{2}\cos ^{4}\theta +\frac{1}{4}\left( \widetilde{\lambda }_{3}+\widetilde{%
\lambda }_{5}\right) \sin ^{2}2\theta -\frac{1}{2}\left( \widetilde{\lambda }%
_{6}\sin ^{2}\theta +\widetilde{\lambda }_{7}\cos ^{2}\theta \right) \sin
2\theta \\
\lambda _{3} &=&\left( \widetilde{\lambda }_{1}+\widetilde{\lambda }_{2}-%
\widetilde{\lambda }_{5}\right) \sin ^{2}2\theta +\widetilde{\lambda }%
_{3}\cos ^{2}2\theta +\frac{1}{2}\left( \widetilde{\lambda }_{7}-\widetilde{%
\lambda }_{6}\right) \sin 4\theta \\
\lambda _{4} &=&\widetilde{\lambda }_{4}
\end{eqnarray*}

\begin{eqnarray*}
\lambda _{5} &=&\frac{1}{2}\left( \widetilde{\lambda }_{1}+\widetilde{%
\lambda }_{2}-\widetilde{\lambda }_{3}\right) \sin ^{2}2\theta +\widetilde{%
\lambda }_{5}\left( \cos ^{4}\theta +\sin ^{4}\theta \right) +\frac{1}{4}%
\left( \widetilde{\lambda }_{7}-\widetilde{\lambda }_{6}\right) \sin 4\theta
\\
\lambda _{6} &=&2\left( \widetilde{\lambda }_{2}\sin ^{2}\theta -\widetilde{%
\lambda }_{1}\cos ^{2}\theta \right) \sin 2\theta +\frac{1}{2}\left( 
\widetilde{\lambda }_{3}+\widetilde{\lambda }_{5}\right) \sin 4\theta \\
&&+\widetilde{\lambda }_{6}\left( 4\cos ^{2}\theta -3\right) \cos ^{2}\theta
+\widetilde{\lambda }_{7}\left( \cos 2\theta +2\cos ^{2}\theta \right) \sin
^{2}\theta \\
\lambda _{7} &=&2\left( \widetilde{\lambda }_{2}\cos ^{2}\theta -\widetilde{%
\lambda }_{1}\sin ^{2}\theta \right) \sin 2\theta -\frac{1}{2}\left( 
\widetilde{\lambda }_{3}+\widetilde{\lambda }_{5}\right) \sin 4\theta \\
&&+\widetilde{\lambda }_{6}\left( 4\cos ^{2}\theta -1\right) \sin ^{2}\theta
+\widetilde{\lambda }_{7}\left( \cos 2\theta -2\sin ^{2}\theta \right) \cos
^{2}\theta
\end{eqnarray*}

\begin{eqnarray}
\lambda _{8} &=&\widetilde{\lambda }_{8}\cos ^{2}\theta +\widetilde{\lambda }%
_{9}\sin ^{2}\theta +\frac{1}{2}\widetilde{\lambda }_{10}\sin 2\theta  \notag
\\
\lambda _{9} &=&\left( \widetilde{\lambda }_{8}\sin ^{2}\theta +\widetilde{%
\lambda }_{9}\cos ^{2}\theta -\frac{1}{2}\widetilde{\lambda }_{10}\sin
2\theta \right)  \notag \\
\lambda _{10} &=&\left[ \left( \widetilde{\lambda }_{9}-\widetilde{\lambda }%
_{8}\right) \sin 2\theta +\widetilde{\lambda }_{10}\cos 2\theta \right]
\label{paramet rotated}
\end{eqnarray}

\subsection{Tadpoles}

From now on, we shall consider the potential with invariance under charge
conjugation \cite{Santos}, i.e. $\mu _{4}=\lambda _{8}=\lambda _{9}=\lambda
_{10}=0$. However, not any other discrete or continuous symmetry is assumed.
In that case the tadpoles are given by 
\begin{equation*}
T=T_{3}h_{1}+T_{7}h_{2}
\end{equation*}%
where%
\begin{eqnarray}
T_{3} &\equiv &-\mu _{1}^{2}v_{1}-\frac{1}{2}\mu _{3}^{2}v_{2}+\lambda
_{1}v_{1}^{3}+\frac{1}{2}\lambda _{3}v_{1}v_{2}^{2}+\frac{1}{2}\lambda
_{5}v_{1}v_{2}^{2}+\frac{3}{4}\lambda _{6}v_{1}^{2}v_{2}+\frac{1}{4}\lambda
_{7}v_{2}^{3}  \notag \\
T_{7} &\equiv &-\mu _{2}^{2}v_{2}-\frac{1}{2}\mu _{3}^{2}v_{1}+\lambda
_{2}v_{2}^{3}+\frac{1}{2}\lambda _{3}v_{1}^{2}v_{2}+\frac{1}{2}\lambda
_{5}v_{1}^{2}v_{2}+\frac{1}{4}\lambda _{6}v_{1}^{3}+\frac{3}{4}\lambda
_{7}v_{2}^{2}v_{1}  \label{tadp no rot}
\end{eqnarray}%
these tadpoles coincide with the minimum conditions, applying $\mu
_{4}=\lambda _{8}=\lambda _{9}=\lambda _{10}=0$. Now, we find the relation
among the tadpoles in both parameterizations by using Eq.(\ref{tadp no rot}%
), and Eqs. (\ref{doubletnotri}, \ref{rotdoublet}).

\begin{eqnarray*}
T_{3}h_{1}+T_{7}h_{2} &=&T_{3}\left( h_{1}^{\prime }\cos \theta
+h_{2}^{\prime }\sin \theta \right) +T_{7}\left( -h_{1}^{\prime }\sin \theta
+\cos \theta h_{2}^{\prime }\right) \\
&=&\left( T_{3}\cos \theta -T_{7}\sin \theta \right) h_{1}^{\prime }+\left(
T_{3}\sin \theta +T_{7}\cos \theta \right) h_{2}^{\prime } \\
&=&T_{3}^{\prime }h_{1}^{\prime }+T_{7}^{\prime }h_{2}^{\prime }
\end{eqnarray*}%
from which we see that the tadpoles in both parameterizations are related
through the rotation 
\begin{equation}
\left( 
\begin{array}{c}
T_{3}^{\prime } \\ 
T_{7}^{\prime }%
\end{array}%
\right) \equiv \left( 
\begin{array}{cc}
\cos \theta & -\sin \theta \\ 
\sin \theta & \cos \theta%
\end{array}%
\right) \left( 
\begin{array}{c}
T_{3} \\ 
T_{7}%
\end{array}%
\right)  \label{tadp relation}
\end{equation}%
As a proof of consistency, we can check that from (\ref{tadp no rot}), and
from (\ref{paramet rotated}) the following relations are obtained after a
bit of algebra 
\begin{eqnarray}
T_{3}\cos \theta -T_{7}\sin \theta &=&-\widetilde{\mu }_{1}^{2}v_{1}^{\prime
}-\frac{1}{2}\widetilde{\mu }_{3}^{2}v_{2}^{\prime }+\widetilde{\lambda }%
_{1}v_{1}^{\prime 3}+\frac{1}{2}\widetilde{\lambda }_{3}v_{1}^{\prime
}v_{2}^{\prime 2}+\frac{1}{2}\widetilde{\lambda }_{5}v_{1}^{\prime
}v_{2}^{\prime 2}+\frac{3}{4}\widetilde{\lambda }_{6}v_{1}^{\prime
2}v_{2}^{\prime }+\frac{1}{4}\widetilde{\lambda }_{7}v_{2}^{\prime 3}  \notag
\\
T_{3}\sin \theta +T_{7}\cos \theta &=&-\widetilde{\mu }_{2}^{2}v_{2}^{\prime
}-\frac{1}{2}\widetilde{\mu }_{3}^{2}v_{1}^{\prime }+\widetilde{\lambda }%
_{2}v_{2}^{\prime 3}+\frac{1}{2}\widetilde{\lambda }_{3}v_{1}^{\prime
2}v_{2}^{\prime }+\frac{1}{2}\widetilde{\lambda }_{5}v_{1}^{\prime
2}v_{2}^{\prime }+\frac{1}{4}\widetilde{\lambda }_{6}v_{1}^{\prime 3}+\frac{3%
}{4}\widetilde{\lambda }_{7}v_{2}^{\prime 2}v_{1}^{\prime }
\label{tadp relat}
\end{eqnarray}%
where 
\begin{equation}
\left( 
\begin{array}{c}
v_{1}^{\prime } \\ 
v_{2}^{\prime }%
\end{array}%
\right) \equiv \left( 
\begin{array}{cc}
\cos \theta & -\sin \theta \\ 
\sin \theta & \cos \theta%
\end{array}%
\right) \left( 
\begin{array}{c}
v_{1} \\ 
v_{2}%
\end{array}%
\right)  \label{VEVrot}
\end{equation}%
relates the VEV's between both parameterizations\footnote{%
Observe that such rotation leaves invariant the quantity $%
v_{1}^{2}+v_{2}^{2}=v_{1}^{\prime 2}+v_{2}^{\prime 2}=\frac{2m_{w}^{2}}{g^{2}%
}$, as it should be.}. From Eqs. (\ref{tadp relation}) and (\ref{tadp relat}%
) we get%
\begin{eqnarray}
T_{3}^{\prime } &=&-\widetilde{\mu }_{1}^{2}v_{1}^{\prime }-\frac{1}{2}%
\widetilde{\mu }_{3}^{2}v_{2}^{\prime }+\widetilde{\lambda }%
_{1}v_{1}^{\prime 3}+\frac{1}{2}\widetilde{\lambda }_{3}v_{1}^{\prime
}v_{2}^{\prime 2}+\frac{1}{2}\widetilde{\lambda }_{5}v_{1}^{\prime
}v_{2}^{\prime 2}+\frac{3}{4}\widetilde{\lambda }_{6}v_{1}^{\prime
2}v_{2}^{\prime }+\frac{1}{4}\widetilde{\lambda }_{7}v_{2}^{\prime 3}  \notag
\\
T_{7}^{\prime } &=&-\widetilde{\mu }_{2}^{2}v_{2}^{\prime }-\frac{1}{2}%
\widetilde{\mu }_{3}^{2}v_{1}^{\prime }+\widetilde{\lambda }%
_{2}v_{2}^{\prime 3}+\frac{1}{2}\widetilde{\lambda }_{3}v_{1}^{\prime
2}v_{2}^{\prime }+\frac{1}{2}\widetilde{\lambda }_{5}v_{1}^{\prime
2}v_{2}^{\prime }+\frac{1}{4}\widetilde{\lambda }_{6}v_{1}^{\prime 3}+\frac{3%
}{4}\widetilde{\lambda }_{7}v_{2}^{\prime 2}v_{1}^{\prime }
\label{tadp relat2}
\end{eqnarray}%
and comparing with Eq. (\ref{tadp no rot}) we see that the form of the
tadpole is preserved by changing the original parameters with the prime
parameters.

Finally, it can be checked that when one of the VEV vanishes, its
corresponding tadpole vanishes as well, this is an important requirement to
preserve the renormalizability of the theory \cite{tadpole}.

\subsection{Higgs boson masses}

Another important proof of consistency is to verify that both
parameterizations predict the same masses for the Higgs bosons. We shall use
once again, the potential with $C-$invariance. In a general
parameterization, the minimal conditions are reduced to 
\begin{eqnarray*}
\mu _{1}v_{1} &=&\left( -\frac{1}{2}\mu _{3}v_{2}+\lambda _{1}v_{1}^{3}+%
\frac{1}{2}\lambda _{3}v_{2}^{2}v_{1}+\frac{1}{2}\lambda _{5}v_{2}^{2}v_{1}+%
\frac{3}{4}\lambda _{6}v_{1}^{2}v_{2}+\frac{1}{4}\lambda _{7}v_{2}^{3}\right)
\\
\mu _{2}v_{2} &=&\left( -\frac{1}{2}\mu _{3}v_{1}+\lambda _{2}v_{2}^{3}+%
\frac{1}{2}\lambda _{3}v_{1}^{2}v_{2}+\frac{1}{2}\lambda _{5}v_{1}^{2}v_{2}+%
\frac{1}{4}\lambda _{6}v_{1}^{3}+\frac{3}{4}\lambda _{7}v_{1}v_{2}^{2}\right)
\end{eqnarray*}%
the mass matrix is obtained by using once again $\mu _{4}=\lambda _{8}$ $%
=\lambda _{9}=\lambda _{10}=0$.\ Let us start with the matrix elements
corresponding to the scalar Higgs bosons $m_{H^{0}},m_{h^{0}}$. If we assume
that both VEV are different from zero and utilize the minimum conditions, we
obtain the following mass matrix 
\begin{equation}
\left( 
\begin{array}{cc}
M_{33}^{2} & M_{37}^{2} \\ 
M_{37}^{2} & M_{77}^{2}%
\end{array}%
\right)  \label{mass37param}
\end{equation}%
with

\begin{eqnarray}
M_{33}^{2} &=&\frac{1}{4v_{1}}\left( 2\mu _{3}^{2}v_{2}+8\lambda
_{1}v_{1}^{3}+3\lambda _{6}v_{1}^{2}v_{2}-\lambda _{7}v_{2}^{3}\allowbreak
\right)  \notag \\
M_{37}^{2} &=&-\allowbreak \frac{1}{2}\mu _{3}^{2}+\frac{3}{4}\lambda
_{7}v_{2}^{2}+\frac{3}{4}\lambda _{6}v_{1}^{2}+\lambda _{3}\allowbreak
v_{1}v_{2}+\lambda _{5}v_{1}v_{2}  \notag \\
M_{77}^{2} &=&\frac{1}{4v_{2}}\left( 2\mu _{3}^{2}v_{1}+8\lambda
_{2}v_{2}^{3}-\lambda _{6}v_{1}^{3}+3\lambda _{7}v_{2}^{2}v_{1}\right)
\label{terms37param}
\end{eqnarray}%
For the sake of simplicity, we just show that the determinant of this matrix
(i.e. the product of the squared masses), coincides for two
parameterizations connected by a transformation like (\ref{rotdoublet}). The
mass matrix in any other parameterization with both VEV different from zero,
have the same form as (\ref{mass37param}, \ref{terms37param}) but replacing $%
\mu _{i}^{2}\rightarrow \widetilde{\mu }_{i}^{2},\;\lambda _{i}\rightarrow 
\widetilde{\lambda }_{i}.\;$It is a fact of cumbersome algebra to demostrate
that 
\begin{equation*}
M_{33}^{2}M_{77}^{2}-\left( M_{37}^{2}\right) ^{2}=\widetilde{M}_{33}^{2}%
\widetilde{M}_{77}^{2}-\left( \widetilde{M}_{37}^{2}\right) ^{2}
\end{equation*}%
this demostration is carried out by taking into account the relations (\ref%
{paramet rotated}) among the parameters in both bases. In a similar fashion,
we can show that the eigenvalues coincide in both bases. Therefore, the
Higgs boson masses are equal in both parameterizations as it must be.
Finally, if the angle of rotation is chosen such that one of the VEV is
zero, (e.g. $v_{2}=0$) in one of the bases, then the minimum conditions and
mass matrix elements become much simpler, and the equality is easier to
demonstrate.

By the same token, we can check that for the other Higgs mass matrices the
determinants and eigenvalues are invariant under the transformation (\ref%
{rotdoublet}). Showing that the observables are not altered by this change
of basis.

\section{VEV with complex phases}

If we assume that one of the VEV in the prime basis (say $v_{2}^{\prime }$)
acquires a complex phase, we can eliminate this VEV by making a complex
rotation. However, even if we start with real parameters in the potential of
the prime parameterization, after the complex rotation some of these
parameters acquire complex values in the new parameterization. In that case
the change of basis has traslated the CP violation sources from the VEV to
the parameters of the potential. In other words the original spontaneous CP
violation has been transformed into an explicit violation of such symmetry.

\section{Conclusions}

The simplest extension of the standard model is the so called two Higgs
doblet model. This is a well motivated model from both theoretical and
phenomenological grounds \cite{Hunter}. For such model, it is customary (in
the absence of additional symmetries) to make a rotation in order to get rid
of one of the vacuum expectation values. The literature has discussed this
rotation in the framework of the Yukawa sector. However, it is important to
check that the physical content of the potential also remains invariant
under the transformation made in the Yukawa sector, especially because a
charge conjugation invariance in the potential is demanded even when no
additional symmetries in the Yukawa sector are demanded. We show by finding
the transformation in the parameters of the potential, that the physical
content of the potential remains invariant even when $C$-invariance is
demanded. Such invariance is demostrated by examining the tadpoles and the
physical spectrum of the Higgs sector, and showing their invariance under
the transformation described above. In particular, it worths emphasizing
that when one of the VEV is null, the corresponding tadpole also vanishes,
this feature is essential to preserve the renormalizability of the theory.

On the other hand, we can realize that the models type I and II have a
remarkable difference with respect to the model type III, since it is well
known that the former two ones are highly dependent on the $\tan \beta \;$%
parameter while the latter is not. We can see the difference from the point
of view of symmetries, the 2HDM is constructed in such a way that we make an
exact \textquotedblleft duplicate\textquotedblright\ of the SM Higgs
doublet. These doublets have the same quantum numbers and are consequently
indistinguishable (at least at this step). Owing to this indistinguibility
we can perform the rotation described above$\ $without any physical
consequences (it is in fact a change of basis). It means that the model is
invariant under a global SO(2) transformation of the \textquotedblleft
bidoublet\textquotedblright\ $\left( 
\begin{array}{cc}
\Phi _{1} & \Phi _{2}%
\end{array}%
\right) ^{T}$. However, it is very common to impose a discrete symmetry on
the Higgs doublets $\left( \Phi _{1}\rightarrow \Phi _{1},\;\Phi
_{2}\rightarrow -\Phi _{2}\right) \ $\cite{Weinberg} or a global $U\left(
1\right) $ symmetry $\left( \Phi _{1}\rightarrow \Phi _{1},\;\Phi
_{2}\rightarrow e^{i\varphi }\Phi _{2}\right) $ \cite{Santos} to prevent
dangerous flavor changing neutral currents or to study the Higgs sector of
some new physics scenarios such as the minimal supersymmetric standard
model. In that case, we are introducing a distinguibility between the
doublets, because they acquire very different couplings to the fermions
(models type I and II). Of course, we could have defined the symmetry in the
opposite way $\left( \Phi _{1}\rightarrow -\Phi _{1},\;\Phi _{2}\rightarrow
\Phi _{2}\right) \;$but once we have chosen one of them, we cannot
interchange $\Phi _{1}\leftrightarrow \Phi _{2}\;$anymore, without changing
the physical content. Such fact breaks explicity the SO(2) symmetry of the
\textquotedblleft bidoublet\textquotedblright . On the other hand, it is
precisely this symmetry what allows us to absorb the $\tan \beta \;$%
parameter, and since models type I and II do not have that symmetry, we are
not able to absorb it properly.

There is another interesting way to see why the rotation can be carried out
when $C-$invariance is imposed while it cannot be applied in the case of
imposing a discrete symmetry. $C-$invariance (in certain basis) requires the
vanishing of the parameters $\mu _{4}^{2},\lambda _{8},\lambda _{9},\lambda
_{10}$; however, Eqs. (\ref{paramet rotated}) shows that such parameters
only transform among themselves i.e. the subset $\mu _{4}^{2},\lambda
_{8},\lambda _{9},\lambda _{10}$ is written in terms of the subset $%
\widetilde{\mu }_{4}^{2},\widetilde{\lambda }_{8},\widetilde{\lambda }_{9},%
\widetilde{\lambda }_{10}$. Such transformations also show that if the
subset in certain basis vanishes, the corresponding subset in any other
basis vanishes as well. By contrast, the discrete symmetry requires the
vanishing of the subset $\mu _{3}^{2},\mu _{4}^{2},\lambda _{6},\lambda
_{7},\lambda _{8},\lambda _{9}$ but this subset do not transform among
themselves as Eqs. (\ref{paramet rotated}) show, from which we see that this
symmetry is not (in general) preserved by the rotation.

Finally, when we consider a phase in a VEV and make a complex rotation that
eliminates such VEV, the change of basis translates the source of CP
violation from the VEV to the parameters of the potential. It means that the
original spontaneous violation of $CP$ has become an explicit violation of
such symmetry when the basis is changed.

\section{Acknowledgements}

We thank to Fundación Banco de la República, and to DINAIN, for their
financial support.

\end{document}